
\font\twelverm=cmr10 scaled 1200    \font\twelvei=cmmi10 scaled 1200
\font\twelvesy=cmsy10 scaled 1200   \font\twelveex=cmex10 scaled 1200
\font\twelvebf=cmbx10 scaled 1200   \font\twelvesl=cmsl10 scaled 1200
\font\twelvett=cmtt10 scaled 1200   \font\twelveit=cmti10 scaled 1200

\skewchar\twelvei='177   \skewchar\twelvesy='60


\def\twelvepoint{\normalbaselineskip=12.4pt
  \abovedisplayskip 12.4pt plus 3pt minus 9pt
  \belowdisplayskip 12.4pt plus 3pt minus 9pt
  \abovedisplayshortskip 0pt plus 3pt
  \belowdisplayshortskip 7.2pt plus 3pt minus 4pt
  \smallskipamount=3.6pt plus1.2pt minus1.2pt
  \medskipamount=7.2pt plus2.4pt minus2.4pt
  \bigskipamount=14.4pt plus4.8pt minus4.8pt
  \def\rm{\fam0\twelverm}          \def\it{\fam\itfam\twelveit}%
  \def\sl{\fam\slfam\twelvesl}     \def\bf{\fam\bffam\twelvebf}%
  \def\mit{\fam 1}                 \def\cal{\fam 2}%
  \def\tt{\twelvett}
  \textfont0=\twelverm   \scriptfont0=\tenrm   \scriptscriptfont0=\sevenrm
  \textfont1=\twelvei    \scriptfont1=\teni    \scriptscriptfont1=\seveni
  \textfont2=\twelvesy   \scriptfont2=\tensy   \scriptscriptfont2=\sevensy
  \textfont3=\twelveex   \scriptfont3=\twelveex  \scriptscriptfont3=\twelveex
  \textfont\itfam=\twelveit
  \textfont\slfam=\twelvesl
  \textfont\bffam=\twelvebf \scriptfont\bffam=\tenbf
  \scriptscriptfont\bffam=\sevenbf
  \normalbaselines\rm}

\def\beginlinemode{\endmode
  \begingroup\parskip=0pt \obeylines\def\endmode{\par\endgroup}}
\def\beginparmode{\endmode
  \begingroup \def\endmode{\par\endgroup}}
\let\endmode=\par
{\obeylines\gdef\
{}}
\def\singlespace{\baselineskip=\normalbaselineskip}
\def\oneandahalfspace{\baselineskip=\normalbaselineskip
  \multiply\baselineskip by 3 \divide\baselineskip by 2}
\def\doublespace{\baselineskip=\normalbaselineskip \multiply\baselineskip by 2}
\newcount\firstpageno
\firstpageno=2
\footline={\ifnum\pageno<\firstpageno{\hfil}\else{\hfil\twelverm\folio\hfil}\fi}
\let\rawfootnote=\footnote            
\def\footnote#1#2{{\rm\singlespace\parindent=0pt\rawfootnote{#1}{#2}}}
\def\raggedcenter{\leftskip=4em plus 12em \rightskip=\leftskip
  \parindent=0pt \parfillskip=0pt \spaceskip=.3333em \xspaceskip=.5em
  \pretolerance=9999 \tolerance=9999
  \hyphenpenalty=9999 \exhyphenpenalty=9999 }
\def\dateline{\centerline{\ifcase\month\or
  January\or February\or March\or April\or May\or June\or
  July\or August\or September\or October\or November\or December\fi
  \space\number\year}}
\hsize=6truein
\vsize=8.5truein

\parskip=\medskipamount
\twelvepoint         
\doublespace         
\overfullrule=0pt    
\def\title                      
  {\null\vskip 3pt plus 0.2fill
   \beginlinemode \doublespace \raggedcenter \bf}

\def\author                     
  {\vskip 3pt plus 0.2fill \beginlinemode
   \singlespace \raggedcenter}

\def\affil                      
  {\vskip 3pt plus 0.1fill \beginlinemode
   \oneandahalfspace \raggedcenter \sl}

\def\abstract                   
  {\vskip 3pt plus 0.3fill \beginparmode
   \doublespace \narrower ABSTRACT: }

\def\endtitlepage               
  {\endpage                     
   \body}

\def\body                       
  {\beginparmode}               

\def\refto#1{$^{#1}$}           

\def\references                 
   {\noindent{\bf REFERENCES}\medskip
   \beginparmode
   \frenchspacing \parindent=0pt \leftskip=0.5truecm
   \parskip=5pt plus 3pt \everypar{\hangindent=\parindent}}

\gdef\refis#1{\indent\hbox to 0pt{\hss#1.~}}    

\def\endreferences{\body}
\def\endpage                    
  {\vfill\eject}

\def\endpaper                   
  {\endmode\vfill\supereject}

\def\ref#1{Ref. #1}                     
\def\Ref#1{Ref. [#1]}                     

\def\cite#1{{#1}}
\def\[#1]{[\cite{#1}]}
\def\(#1){(\call{#1})}
\def\call#1{{#1}}\def\taghead#1{{#1}}
\def\frac#1#2{{\textstyle{#1 \over #2}}}

\def\sla{\raise.15ex\hbox{$/$}\kern-.57em}
\def\leaderfill{\leaders\hbox to 1em{\hss.\hss}\hfill}
\def\twiddle{\lower.9ex\rlap{$\kern-.1em\scriptstyle\sim$}}
\def\bigtwiddle{\lower1.ex\rlap{$\sim$}}
\def\gtwid{\mathrel{\raise.3ex\hbox{$>$\kern-.75em\lower1ex\hbox{$\sim$}}}}
\def\ltwid{\mathrel{\raise.3ex\hbox{$<$\kern-.75em\lower1ex\hbox{$\sim$}}}}
\def\square{\kern1pt\vbox{\hrule height 1.2pt\hbox{\vrule width 1.2pt\hskip 3pt
   \vbox{\vskip 6pt}\hskip 3pt\vrule width 0.6pt}\hrule height 0.6pt}\kern1pt}

\catcode`@=11
\newcount\r@fcount \r@fcount=0
\newcount\r@fcurr
\immediate\newwrite\reffile
\newif\ifr@ffile\r@ffilefalse
\def\w@rnwrite#1{\ifr@ffile\immediate\write\reffile{#1}\fi\message{#1}}

\def\writer@f#1>>{}
\def\referencefile{
\r@ffiletrue\immediate\openout\reffile=\jobname.ref%
  \def\writer@f##1>>{\ifr@ffile\immediate\write\reffile%
    {\noexpand\refis{##1} = \csname r@fnum##1\endcsname = %
     \expandafter\expandafter\expandafter\strip@t\expandafter%
     \meaning\csname r@ftext\csname r@fnum##1\endcsname\endcsname}\fi}%
  \def\strip@t##1>>{}}

\def\citeall#1{\xdef#1##1{#1{\noexpand\cite{##1}}}}
\def\cite#1{\each@rg\citer@nge{#1}}

\def\each@rg#1#2{{\let\thecsname=#1\expandafter\first@rg#2,\end,}}
\def\first@rg#1,{\thecsname{#1}\apply@rg}
\def\apply@rg#1,{\ifx\end#1\let\next=\relax%
\else,\thecsname{#1}\let\next=\apply@rg\fi\next}%

\def\citer@nge#1{\citedor@nge#1-\end-}
\def\citer@ngeat#1\end-{#1}
\def\citedor@nge#1-#2-{\ifx\end#2\r@featspace#1
  \else\citel@@p{#1}{#2}\citer@ngeat\fi}
\def\citel@@p#1#2{\ifnum#1>#2{\errmessage{Reference range #1-#2\space is bad.}
    \errhelp{If you cite a series of references by the notation M-N, then M and
    N must be integers, and N must be greater than or equal to M.}}\else%
{\count0=#1\count1=#2\advance\count1
by1\relax\expandafter\r@fcite\the\count0,%
  \loop\advance\count0 by1\relax
    \ifnum\count0<\count1,\expandafter\r@fcite\the\count0,%
  \repeat}\fi}

\def\r@featspace#1#2 {\r@fcite#1#2,}
\def\r@fcite#1,{\ifuncit@d{#1}
    \newr@f{#1}
    \expandafter\gdef\csname r@ftext\number\r@fcount\endcsname%
                      {\message{Reference #1 to be supplied.}
                       \writer@f#1>>#1 to be supplied.\par}
\fi
\csname r@fnum#1\endcsname}
\def\ifuncit@d#1{\expandafter\ifx\csname r@fnum#1\endcsname\relax}%
\def\newr@f#1{\global\advance\r@fcount by1%
      \expandafter\xdef\csname r@fnum#1\endcsname{\number\r@fcount}}

\let\r@fis=\refis
\def\refis#1#2#3\par{\ifuncit@d{#1}%

    \w@rnwrite{Reference #1=\number\r@fcount\space is not cited up to now.}\fi%
  \expandafter\gdef\csname r@ftext\csname r@fnum#1\endcsname\endcsname%
  {\writer@f#1>>#2#3\par}}

\def\ignoreuncited{
   \def\refis##1##2##3\par{\ifuncit@d{##1}%
     \else\expandafter\gdef\csname r@ftext\csname
r@fnum##1\endcsname\endcsname%
     {\writer@f##1>>##2##3\par}\fi}}

\def\r@ferr{\endreferences\errmessage{I was expecting to see
\noexpand\endreferences before now;  I have inserted it here.}}
\let\r@ferences=\references
\def\references{\r@ferences\def\endmode{\r@ferr\par\endgroup}}

\let\endr@ferences=\endreferences
\def\endreferences{\r@fcurr=0
   {\loop\ifnum\r@fcurr<\r@fcount
    \advance\r@fcurr by 1\relax\expandafter\r@fis\expandafter{\number\r@fcurr}%
    \csname r@ftext\number\r@fcurr\endcsname%
  \repeat}\gdef\r@ferr{}\endr@ferences}
\let\r@fend=\endpaper\gdef\endpaper{\ifr@ffile
\immediate\write16{Cross References written on []\jobname.REF.}\fi\r@fend}
\def\reftorange#1#2#3{[{\cite{#1}-\setbox0=\hbox{\cite{#2}}\cite{#3}}]}

\catcode`@=12
\citeall\refto\citeall\ref\citeall\Ref
\catcode`@=11
\newcount\tagnumber\tagnumber=0
\immediate\newwrite\eqnfile\newif\if@qnfile\@qnfilefalse
\def\write@qn#1{}\def\writenew@qn#1{}
\def\w@rnwrite#1{\write@qn{#1}\message{#1}}
\def\@rrwrite#1{\write@qn{#1}\errmessage{#1}}
\def\taghead#1{\gdef\t@ghead{#1}\global\tagnumber=0}
\def\t@ghead{}\expandafter\def\csname @qnnum-3\endcsname
  {{\t@ghead\advance\tagnumber by -3\relax\number\tagnumber}}
\expandafter\def\csname @qnnum-2\endcsname
  {{\t@ghead\advance\tagnumber by -2\relax\number\tagnumber}}
\expandafter\def\csname @qnnum-1\endcsname
  {{\t@ghead\advance\tagnumber by -1\relax\number\tagnumber}}
\expandafter\def\csname @qnnum0\endcsname
  {\t@ghead\number\tagnumber}
\expandafter\def\csname @qnnum+1\endcsname
  {{\t@ghead\advance\tagnumber by 1\relax\number\tagnumber}}
\expandafter\def\csname @qnnum+2\endcsname
  {{\t@ghead\advance\tagnumber by 2\relax\number\tagnumber}}
\expandafter\def\csname @qnnum+3\endcsname
  {{\t@ghead\advance\tagnumber by 3\relax\number\tagnumber}}
\def\equationfile{\@qnfiletrue\immediate\openout\eqnfile=\jobname.eqn%
  \def\write@qn##1{\if@qnfile\immediate\write\eqnfile{##1}\fi}
  \def\writenew@qn##1{\if@qnfile\immediate\write\eqnfile
    {\noexpand\tag{##1} = (\t@ghead\number\tagnumber)}\fi}}
\def\callall#1{\xdef#1##1{#1{\noexpand\call{##1}}}}
\def\call#1{\each@rg\callr@nge{#1}}
\def\each@rg#1#2{{\let\thecsname=#1\expandafter\first@rg#2,\end,}}
\def\first@rg#1,{\thecsname{#1}\apply@rg}
\def\apply@rg#1,{\ifx\end#1\let\next=\relax%
\else,\thecsname{#1}\let\next=\apply@rg\fi\next}
\def\callr@nge#1{\calldor@nge#1-\end-}\def\callr@ngeat#1\end-{#1}
\def\calldor@nge#1-#2-{\ifx\end#2\@qneatspace#1 %
  \else\calll@@p{#1}{#2}\callr@ngeat\fi}
\def\calll@@p#1#2{\ifnum#1>#2{\@rrwrite{Equation range #1-#2\space is bad.}
\errhelp{If you call a series of equations by the notation M-N, then M and
N must be integers, and N must be greater than or equal to M.}}\else%
{\count0=#1\count1=#2\advance\count1 by1\relax\expandafter\@qncall\the\count0,%
  \loop\advance\count0 by1\relax%
    \ifnum\count0<\count1,\expandafter\@qncall\the\count0,  \repeat}\fi}
\def\@qneatspace#1#2 {\@qncall#1#2,}
\def\@qncall#1,{\ifunc@lled{#1}{\def\next{#1}\ifx\next\empty\else
  \w@rnwrite{Equation number \noexpand\(>>#1<<) has not been defined yet.}
  >>#1<<\fi}\else\csname @qnnum#1\endcsname\fi}
\let\eqnono=\eqno\def\eqno(#1){\tag#1}\def\tag#1$${\eqnono(\displayt@g#1 )$$}
\def\aligntag#1\endaligntag  $${\gdef\tag##1\\{&(##1 )\cr}\eqalignno{#1\\}$$
  \gdef\tag##1$${\eqnono(\displayt@g##1 )$$}}
\def\eqalignno#1{\displ@y \tabskip\centering
  \halign to\displaywidth{\hfil$\displaystyle{##}$\tabskip\z@skip
    &$\displaystyle{{}##}$\hfil\tabskip\centering
    &\llap{$\displayt@gpar##$}\tabskip\z@skip\crcr
    #1\crcr}}
\def\displayt@gpar(#1){(\displayt@g#1 )}
\def\displayt@g#1 {\rm\ifunc@lled{#1}\global\advance\tagnumber by1
        {\def\next{#1}\ifx\next\empty\else\expandafter
        \xdef\csname @qnnum#1\endcsname{\t@ghead\number\tagnumber}\fi}%
  \writenew@qn{#1}\t@ghead\number\tagnumber\else
        {\edef\next{\t@ghead\number\tagnumber}%
        \expandafter\ifx\csname @qnnum#1\endcsname\next\else
        \w@rnwrite{Equation \noexpand\tag{#1} is a duplicate number.}\fi}%
  \csname @qnnum#1\endcsname\fi}
\def\ifunc@lled#1{\expandafter\ifx\csname @qnnum#1\endcsname\relax}
\let\@qnend=\end\gdef\end{\if@qnfile
\immediate\write16{Equation numbers written on []\jobname.EQN.}\fi\@qnend}
\catcode`@=12

\ignoreuncited
\baselineskip=24pt
\bigskip
\title{DECOHERENCE, CHAOS, AND THE SECOND LAW}
\bigskip
\author{Wojciech Hubert Zurek and Juan Pablo Paz}
\bigskip
\affil{Theoretical Astrophysics, T-6, Mail Stop B288,}
\vskip -.5truecm
\affil{Los Alamos National Laboratory. Los Alamos, NM 87545}
\bigskip
\bigskip
\abstract{
We investigate implications of decoherence for quantum systems which
are classically chaotic. We show that, in open systems, the rate of von Neumann
entropy
production quickly reaches an asymptotic value which is: (i) independent of the
system-environment coupling, (ii)
dictated by the dynamics of the system, and (iii)
dominated by the largest Lyapunov
exponent. These results shed a new light on the correspondence between quantum
and
classical dynamics as well as on the origins of the ``arrow of time.''
}
\endtitlepage
\body
The relation between classical and quantum chaos has been always somewhat
unclear \refto{Chirikov} and, at times, even strained \refto{Ford}. The cause
of the
difficulties can
be traced to the fact that the defining characteristic of classical chaos
-- {\it sensitive dependence on initial conditions} -- has no quantum
counterpart: it
is defined through
the behavior of neighboring trajectories \refto{Hamiltonian},
a concept which is essentially alien to quantum mechanics.
Moreover, when the natural language of quantum
mechanics of closed systems is adopted, an analogue of the exponential
divergence cannot be found. This is not to deny that many interesting insights
into quantum mechanics have been arrived at by studying quantized versions of
classically chaotic
systems \refto{QuantumCh}. These insights have typically much to do with
the energy spectra, and leave the issue of the relationship between the
quantum and the classical largely open.

The aim of this paper is to investigate implications of the process of
decoherence for quantum chaos. Decoherence is caused by the loss of
phase coherence between the set of preferred quantum states in the Hilbert
space
of the system due to the interaction with the environment \refto{Decoherence}.
Preferred states are singled out by their stability (measured, for example, by
the rate of predictability loss -- the rate of entropy increase) under the
joint influence of the environment and the
self--hamiltonian \refto{ZurekPT}. Thus, the strength and nature of
the coupling with the environment play a crucial role in selecting
preferred states, which -- given the distance--dependent nature of
typical interactions -- explains the special function of the position
observable \refto{Decoherence,ZurekPT}. Coupling with the environment
also sets the decoherence timescale -- the
time on which quantum interference between preferred
states disappear \reftorange{Decoherence}{ZurekPT, PazHZ, ZurekHP}{Zurek86}.
Classicality is then an emergent property of an open quantum
system. It is caused by the incessant monitoring by the environment, the state
of which keeps a ``running record" of the preferred observables of the
evolving quantum
system. For simple quantum systems the programme sketched above
can be carried out rigorously, and yields intuitively appealing results
\refto{Decoherence,Zurek86}.
For example, preferred states of an underdamped harmonic oscillator turn out
to be its coherent states \refto{ZurekHP}.

If decoherence does induce a transition from quantum to classical,
then it should be possible to utilize it in the context of quantum chaos
to establish a more straightforward correspondence between the
behavior of classically chaotic systems and their quantum counterparts.
With this goal in mind we will
consider a classically chaotic system, characterized by a
potential $V(x)$, coupled to an external environment.
A master equation for the density operator of an open quantum system can be
derived under a variety of reasonable assumptions \refto{QBM}.
Here, we shall focus our
considerations on the simplest special case, the high temperature limit
of an ohmic environment, leaving the discussion of other cases (low
temperature regime, non--ohmic environments, etc) for later publications.
In this case, the Wigner function of the system evolves according to
\refto{QBM}:
$$
\dot W \ = \{H,W\}_{PB} + \sum_{n\geq 1}
{\hbar^{2n}(-1)^{n}\over{2^{2n}(2n+1)!}} \partial_x^{2n+1}V\partial_p^{2n+1}W
+ 2 \gamma \partial_p (p W) ~ + D~
\partial^2_{pp} W\eqno(wigner)
$$
where $\gamma$ is the relaxation rate, the
diffusion coefficient is
$D= 2\gamma m k_B T $ ($T$ is the temperature of the
environment and $m$ is the mass of the system).
The first term is the Poisson bracket, which generates the ordinary
Liouville flow. Both the Poisson bracket and the higher derivative terms
result from an expansion of the Moyal bracket,
$\{H,W\}_{MB}=-i\sin(i\hbar\{H,W\}_{PB})/\hbar$,
which generates evolution in phase space of a closed system (this expansion
is valid when $V(x)$ is an analytic function).
The last two terms in \(wigner) arise due to the interaction with the
environment. The first of them produces
relaxation -- gradual loss of energy to the reservoir -- and
the last one diffusion (this diffusion term
is responsible for the {\it decoherence} process).

As is clear from \(wigner), as a consequence of the quantum correction
terms that contain higher derivatives,
the Wigner function of an isolated non--linear system does not follow
the classical Liouville flow.
In a classically chaotic system these non--classical
corrections rapidly gain importance, as can be seen by the following argument:
When a chaotic flow is investigated locally in the
phase space, the evolution operator can be expanded in coordinates
``comoving" with a reference
trajectory. The pattern of flow of the neighboring
trajectories is then generated by the Jacobian of the  transformation.
Eigenvalues of this Jacobian are known as {\it local Lyapunov exponents}
$\lambda_i$, which must sum to zero, since the transformation preserves phase
space volume. Eigenvectors define directions in the phase space along which
the neighboring trajectories either only expand ($\lambda_i > 0$) or only
contract ($\lambda_i < 0$) with respect to the fiducial trajectory at a rate
given by the corresponding Lyapunov exponent \refto{Hamiltonian}.
The exponential contraction rapidly generates small scale
structure in the Wigner function. Thus, the high derivative terms are
$\partial_p^n W\propto \sigma_p^{-n}W$ with $\sigma_p\propto
\sigma_p(0)\exp(\lambda t)$ where $\lambda$ is a Lyapunov exponent. Hence,
non
classical corrections will become important after a characteristic
crossover time $t_\chi$ which
can be estimated by comparing the magnitude of the nonlinear corrections with
the contribution of the Poisson bracket in equation \(wigner).
Defining a characteristic length for the non--linear
terms in the potential as $\chi_n\propto(\partial_xV/\partial_x^{n+1}V)^{1/n}$
we obtain that the n--th order term in \(wigner) becomes comparable with
the Poisson bracket at a time $t_\chi^{(n)}$ given by
$$
t_\chi^{(n)}\propto\lambda^{-1}\log(\chi_n\sigma_p(0)/\hbar)\eqno(tchi)
$$

One of the points we want to make here is that, in the presence
of decoherence, the regime in which the quantum
correction become important is easily avoided. The main reason for
this is the existence of diffusive effects which put a lower bound on
the small scale structure which can be produced by the chaotic evolution.
As a result, $\sigma_p(t)$ can never become sufficiently small to result in
large corrections: Poisson bracket is an excellent approximation of the
quantum Moyal bracket for smooth Wigner functions.
This line of reasoning, as we will
argue below, has also important consequences concerning the rate of
entropy production.

To simplify our analysis, based on equation \(wigner),
we will neglect the relaxation term which, as pointed out in the
literature \refto{Zurek86},
can be made arbitrarily small without decreasing the effectiveness
of the decoherence process
(e.g. by letting $\gamma$ approach zero while keeping $D$ constant).
In this way, we will focus in the important
{\it reversible classical limit} \refto{ZurekPT,Zurek86,GMH}.
We will not limit ourselves to
models leading to Eq. \(wigner) which, as they break the symmetry between
$x$ and $p$ coupling with the environment through
position, have momentum diffusion only. It is convenient
(especially in the context of quantum optics, where the ``rotating
wave approximation" can be invoked) to use a symmetric coupling
(of the form $a^{\dagger}b + a b^{\dagger}$, where $a$ and $b$ are the
annihilation operators of the system and the mode of
the environment \refto{Louisell}). The corresponding
equation differs from \(wigner) in the form of the diffusion  which is
now symmetric, $\propto D (\partial^2_{pp} + \partial^2_{xx})W$.
We shall alternate between using this symmetric diffusion and
the more exact diffusion operator of equation \(wigner)
in the discussion below.

Our objective here is to study the interplay between the evolution
which classically results in an exponential divergence of neighboring
trajectories
(the characteristic feature of chaos) and the destruction of quantum coherence
between a preferred set of states in the Hilbert space of an open system
(a defining feature of decoherence). We shall do this by
using a simple unstable system that still captures the essential features we
want to consider.
In general, a chaotic
geodesic flow pattern is locally analogous to the one occurring near
a saddle point, with stable and unstable directions defined in an obvious
manner. The simplest example of such
a saddle point is afforded by an {\it unstable harmonic oscillator}.
We shall use it as a ``generic" model of locally chaotic phase space dynamics
in our considerations below. In this case, the potential is
$V(x)=-\lambda x^2/2$ ($\lambda$ is the Lyapunov exponent)
and equation \(wigner) reduces to a very simple form.
We will use the lessons learned from this simple unstable system to
argue that quantum corrections, which in the case of
the unstable oscillator vanish identically, can be neglected whenever
decoherence is effective.
Our analysis will also show that three different
stages of the evolution can be identified:
(i) decoherence, (ii)
approximately reversible Liouville flow and (iii) irreversible
diffusion--dominated evolution.

To analyze in detail the unstable oscillator
it is convenient to use
contracting and expanding coordinates defined by
${u\atop v}=(p\mp m\lambda x)$.
Evolution generated by Eq. \(wigner) causes exponential
expansion in $v$ and, without the diffusive term, it would
also cause an exponential
contraction in $u$, so that the volume in the phase space (as well as entropy)
would be constant (see Fig. 1). Expansion in $v$ would also result in
an exponential decrease of gradients in that direction. Thus, after a
sufficient number of e-foldings the equation governing evolution of $W$
would be dominated by the expression:
$$ \dot W \ = \ \lambda\Bigl(u \partial_u - v \partial_v +
{1\over 2}\sigma_c^2 \partial_{uu}\Bigr) W \ . \eqno(5)$$
The characteristic dispersion, which will play an important role
below, is
$$\sigma_c^2={2D\over\lambda}.\eqno(C)$$

We can now easily examine the fate of a generic initial quantum state.
The general solution of equation \(5) can be found by noticing that
the eigenfunctions of the operator appearing in its right hand side are
$v^n F_m(u/\sigma_c)$  where $F_m(x)=\exp(-x^2/2) H_{m-1}(x/\sqrt 2)$ and
$H_m(x)$ are Hermite polynomials. Expanding $W(u,v,t)$ in terms of these
eigenfunctions (whose eigenvalue is simply $-(n+m)$) we obtain
$$
W(u,v,t)=\sum_{{n\ge 0}\atop{m\ge 1}} a_{nm} (v
e^{-\lambda t})^n F_m(u) e^{-m\omega t}
\eqno(6)
$$
{}From this expression we see that the Wigner function depends on $v$ only
through
the combination $v_0=v\exp(-\lambda t)$, which is the comoving coordinate.
That is, along this direction, the Wigner function just expands.  Moreover,
after a few dynamical times the most important contribution to
\(6) will always come from the $m=1$ term. Thus, in the contracting direction
the Wigner function approaches a Gaussian with a critical width
and has the form:
$$
W(u,v,t)\approx {1\over{\sqrt{2\pi\sigma^2_c}}} \
e^{-u^2/2\sigma_c^2} \ e^{-\omega t} \int_{-\infty}^\infty du \
W(u,v_0 , t=0).\eqno(7)
$$
The existence of the critical width
$\sigma_c$ is a
consequence of the interplay between the exponential divergence of
trajectories and diffusion.
In effect, a competition between the chaotic evolution (which attempts to
``squeeze" the wavepacket in the contracting direction), and the diffusion
(which has the opposite tendency) leads to a compromise steady state which
results in the Gaussian written above.

How do these general considerations imply the existence of
the three distinguishable phases of the evolution?
The analysis of {\it decoherence},
responsible for the quantum to classical transition, follows simply.
Non--classical states possessing a rapidly oscillating
non--positive $W$ quickly evolve towards
a mixture of localized states
eventually resulting in a positive Wigner function.
For example, if the initial state is a superposition of
two coherent states separated by a distance $L$ (along $u$),
the ratio between the wavelength of the interference fringes
${\ell\propto 1/L}$ and
$\sigma_c$ is ${\sigma_c^2/{\ell^2}}=(\gamma/\lambda)({L/{\lambda_{dB}}})^2$
where $\lambda_{dB}$ is the thermal de Broglie wavelength \refto{Zurek86}.
Therefore, the decoherence time is
$$
\tau_{dec}=\gamma^{-1}\bigl({\lambda_{dB}\over L})^2\eqno(tdec2)
$$
which, for macroscopic scales, is much smaller than the dynamical times
even for very weakly dissipative systems.
Our analysis of the decoherence period is still incomplete since equation \(7)
suggests that the negativity of the Wigner function may persist along the
expanding direction
(in that equation
the initial state is not changed along $v$ but just
``stretched" by the geodesic flow). However,
equation \(5) was obtained by neglecting gradients along $v$.
When a diffusion term $D\partial^2_{vv}W$ is added to the right hand side of
\(5), the eigenfunctions change in such a way that the powers
$(v e^{-\lambda t})^n$ are replaced by Hermite polynomials
$H_n(v/\sqrt 2\sigma_c) e^{-n \lambda t}$.
If we reexpress the solution in terms of
$v_0$ we notice that, for times of the order of $1/\lambda$ only the highest
power in $H_n(v/\sqrt 2\sigma_c)$ survives. This implies that the asymptotic
form of $W$ is no longer given by  \(7), which just contains
the initial state expanded along the unstable
direction. The correct expression
simply contains a smoothed version of the initial condition in which the
details smaller than $\sigma_c$ are washed out along the unstable direction.
Again, oscillations with wavelength $\ell\propto 1/L$ are typically destroyed
after
the decoherence time \(tdec2).

The analysis of the
reversible and irreversible stages can be illustrated by following the
evolution of
a Gaussian $W$. Here, the
existence of $\sigma_c$ is again very important.
The von Neumann entropy ${\cal H}$ of a Gaussian state can be easily related
to the area $A$ enclosed by a $1$--$\sigma$ contour of the Wigner function.
Thus, ${\cal H}$ is a monotonic function of $A$ which, when $A\gg h=2\pi\hbar$
approaches ${\cal H}\simeq k_B\ln A/h$. Using the above equations, one
can show that the rate of entropy production is
$$
\dot {\cal H} = {\dot A\over A}= \lambda\
{\sigma_c^2\over{\sigma_p^2(t)}}\eqno(9)
$$
where $\sigma_p(t)$ is the width of the Gaussian along the direction of
$p$. Consequently  when the width of the Gaussian approaches
the critical value, the entropy growth becomes equal to the positive Lyapunov
exponent $\lambda$.
The evolution of $\dot {\cal H}$ can be approximately analyzed as follows:
one can use \(wigner) to show that
the ratio $R\equiv(\sigma_p(t)/\sigma_c)^2$ evolves according to the equation
$\dot R=2\lambda(1-\beta R)$ where $\beta(t)$, which is related to the degree
of squeezing of
the state, approaches unity exponentially fast. Solving this equation
approximately (using $\beta=1$) one gets:
$$
\dot{\cal H} = \lambda\
\Bigl(1+({\sigma_p^2(0)\over{\sigma_c^2}}-1)\exp(-2\lambda t)\Bigr)^{-1}
\eqno(entro)
$$

Evolution of a classical distribution (corresponding to a
non-negative $W$, initially
smooth on a scale much larger than $\sigma_c$ and
spread over a regular patch with $A\gg h$)
will typically proceed in two different stages (see Fig. 2). The first stage
will be
approximately area--preserving with the evolution dominated by the Liouville
operator. It will last
as long as each of the dimensions of the patch is much larger than the
critical width. During this stage diffusion does little to alter the form of
$W$. The Wigner function is merely ``stretched" or
``contracted" by the geodesic flow so that, with respect to the co-moving
coordinates,
``nothing happens" to $W$. By contrast, when the dimension of the patch
becomes comparable with $\sigma_c$, diffusion will begin to dominate. Further
contraction
will be halted at $\sigma_c$ but the stretching will proceed at the rate
set by the
positive Lyapunov exponent. As a result, the area (or, more generally, the
volume) in phase
space will increase at the rate set by \(entro) with $\sigma_p=\sigma_c$. Using
our approximate
equation \(entro) one can estimate the
time corresponding to the transition from reversible to irreversible evolution:
$$
\tau_c=\lambda^{-1}\ln\bigl({\sigma_p(0)\over{\sigma_c}}
\bigr)\eqno(timescale)
$$

An important condition must be valid in order
to apply the above arguments, illustrated here in the simplest example,
to more complicated non--linear systems: we need to assure that the Wigner
function follows approximately the evolution generated by the
Poisson bracket (i.e. that the effect of the higher derivative terms in
\(wigner) is
small).  As we pointed out above, in the
absence of decoherence the quantum corrections to the evolution of $W$ become
important at a crossover time $t_\chi$ given by equation \(tchi).
These quantum corrections will remain small if diffusion is strong
enough to prevent the formation of small structure in $W$. As we argued above,
the formation of small structure induced by the exponential contraction is
stopped
after a time $\tau_c$. Therefore, the
condition for the Wigner function to evolve classically is
$$t_\chi\gg\tau_c,\eqno(condtimes)
$$
which, taking into account equations \(tchi) and \(timescale), can
be rewritten in the following suggestive way:
$$
\chi_n\sigma_c\gg\hbar.\eqno(condition)
$$
This condition -- a key criterion to assure the correspondence between
quantum and classical dynamics -- assures that $W$ follows the Liouville flow
(albeit with diffusive contributions) and allows
us to apply the conclusions of our previous analysis.


We have demonstrated that chaotic quantum systems can exhibit, in addition to
the very
rapid onset of decoherence, a nearly
reversible phase of evolution which is necessarily followed by an irreversible
stage in which the
entropy increases linearly at the rate determined by the Lyapunov exponents.
By contrast, open quantum systems with regular
classical analogs continue to evolve with little entropy
production (although possibly with a significant change in dynamics
\refto{Milburn}).
This nearly constant rate of (von Neumann) entropy production,
a consequence of the interplay between the chaotic dynamics of
the system and its interaction with the environment,
suggests not only a clear distinction between
the integrable and chaotic systems, but also shows that increase of
entropy in the context of quantum measurement \refto{vonNeumann}
and the dynamical aspects
of the second law are intimately related and can be traced to the same
cause: Impossibility of isolating macroscopic
systems from their environments.

\references

\refis{Chirikov} B.V. Chirikov, Phys. Rep. {\bf 52}, 263 (1979); G. M.
Zaslavsky, Phys. Rep.
{\bf 80}, 157 (1981); M--J. Giannoni, A. Voros, J. Zinn--Justin, eds., {\it
Chaos and Quantum
Physics}, Les Houches Lectures LII (North Holland, Amsterdam, 1991).

\refis{Ford} J. Ford and G. Mantica, Am. J. Phys. {\bf 60}, 1086 (1992).

\refis{Hamiltonian} See, e.g. G. Ioos, R. H. G. Hellmann and R. Stora eds.,
{\it Chaotic Behavior in
Deterministic Systems}, Les Houches Lectures XXXVI (North Holland, Amsterdam,
1983);
R. S. Mc Kay and J. O. Meiss eds., {\it Hamiltonian Dynamical Systems} (Hilger,
Philadelphia, 1987).

\refis{QuantumCh} M. B. Berry, Proc. Roy. Soc. London, {\bf A423}, 219 (1989);
M. Gutzwiller,
{\it Chaos in Classical and Quantum Mechanics} (Springer Verlag, New York,
1990);
F. Haake, {\it Quantum Signature of Chaos} (Springer Verlag, New York, 1990)
and references therein.

\refis{Decoherence} W. H. Zurek, Phys. Rev. {\bf D24}, 1516 (1981); {\it ibid},
{\bf D26}, 1862 (1982); E. Joos and H. D. Zeh, Zeits. Phys. {\bf B59}, 229
(1985).

\refis{ZurekPT} W. H. Zurek, Physics Today {\bf 44}, 36 (1991); {\it ibid},
{\bf 46}, 81 (1993); Prog.
Theor. Phys. {\bf 89}, 281 (1993).

\refis{PazHZ} J. P. Paz, S. Habib and W. H. Zurek, Phys. Rev. {\bf D47}, 488
(1993).

\refis{ZurekHP} W. H. Zurek, S. Habib and J. P. Paz, Phys. Rev. Lett. {70},
1187 (1993).

\refis{Zurek86} W. H. Zurek, in {\it Frontiers of Nonequilibrium Statistical
Mechanics}, G.
T. Moore and M. O. Scully eds., (Plenum, New York, 1986).

\refis{GMH} M. Gell--Mann and J. B. Hartle, Phys. Rev. {\bf D47}, 3345 (1993).

\refis{QBM} A. O. Caldeira and A. J. Leggett, Physica {\bf 121A}, 587 (1983);
W.
G. Unruh and W. H. Zurek, Phys Rev. {\bf D40}, 1071 (1989); B. L. Hu, J. P. Paz
and Y. Zhang, Phys. Rev. {\bf D45}, 2843 (1992); {\it ibid} {\bf D47}, 1576
(1993).

\refis{Louisell} W. H. Louisell, {\it Quantum Statistical Properties of
Radiation} (Wiley, New
York, 1973); V. Buzek and P. Knight; {\it Quantum interference, superposition
states of light
and nonclassical effects}, Prog. in Opt. (1993), to appear.

\refis{vonNeumann} J. von Neumann, {\it Mathematical Foundations of Quantum
Mechanics},
english translation by R. T. Beyer (Princeton Univ. Press, Princeton, 1955); H.
D. Zeh,
{\it The Direction of Time}, (Springer--Verlag, New York, 1991).

\refis{Blatt} J. M. Blatt expressed a similar suspicion in a compelling,
although less rigorous
and completely classical paper, Progr. Theor. Phys. {\bf 22}, 745 (1959)

\refis{Milburn} G.J. Milburn and C.A. Holmes, Phys. Rev. Lett. {\bf 56}, 2237
(1986).

\refis{Casati} G. Cassati, B. V. Chirikov, F. M. Izraileev and J. Ford, {\it
Lectures Notes in
Physics} {\bf 93}, (Springer--Verlag, New York, 1979).

\refis{Dittrich} T. Dittrich and R. Graham, Ann. Phys. (NY) {\bf 200}, 363
(1990), and
references therein.

\refis{Wigner} M. Hillery, R.F. O'Connell, M.O. Scully and E. Wigner, Phys.
Rep. {\bf 106} (1984), 121

\endreferences
\vfill\eject
\centerline{\bf Figure Captions}

\noindent{\bf Figure 1:} Three stages of the evolution of an initially
Gaussian Wigner function
shown in two different coordinate systems. Fig. (a)--(c) display $W$ in
physical coordinates
($u,v$, with $v$ in logarithmic scale). For an open system which evolves
according to equation \(wigner) with $V=\lambda x^2/2$
(forefront), the width of the distribution
reaches the asymptotic value $\sigma_c$. By contrast, when $W$
evolves unitarily
(densely hatched and shown in the back), it continues to be squeezed in $u$.
Fig. (a')--(c') shows the
same three stages of the evolution, but now in co--moving coordinates
($\tilde v= v \exp(\lambda t),
\tilde u= u \exp(-\lambda t)$). In the unitary case
(cross--hatched, shown in the back)
$W$ does not change. The interaction with the environment causes an exponential
increase in the apparent width of $W$ (forefront).
Since the width of the Gaussians in the expanding direction
is approximately the same, the asymptotic regime of the diffusive
evolution leads to an exponential increase of
the area enclosed in $1$--$\sigma$ contour.
Consequently, the entropy increases linearly at a rate
determined by the Lyapunov exponent.

\noindent{\bf Figure 2:} The rate of von Neumann entropy production for the
quantum open
system. The initial state is a Gaussian for which ${\cal H}(t=0)\gg 1$
and with the initial width
along the contracting direction much larger than $\sigma_c$. The (nearly)
reversible and irreversible stages of the evolution are clearly distinguished.

\end